# WHAT IS INTRUMENTALITY IN NEW DIGITAL MUSICAL DEVICES ? A CONTRIBUTION FROM COGNITIVE LINGUISTICS & PSYCHOLOGY

**Cance, Caroline,** Groupe LCPE, Equipe LAM, Institut d'Alembert – UPMC/CNRS/MCC *cance@lam.jussieu.fr*
**Genevois, Hugues,** Equipe LAM, Institut d'Alembert – UPMC/CNRS/MCC *genevois@lam.jussieu.fr*
**Dubois, Danièle,** Groupe LCPE, Equipe LAM, Institut d'Alembert – UPMC/CNRS/MCC *daniele.dubois@upmc.fr*

## RÉSUMÉ

**Etat de l'art en Informatique Musicale**

En inventant de nouveaux dispositifs de production musicale, le XX[e] siècle nous invite à revisiter le concept d'instrument de musique. De nombreux travaux, en musicologie [1], comme en informatique musicale [4, 5, 21] décrivent l'histoire de ces évolutions, les conditions techniques de leur apparition, et les formes originales « d'instrumentalité » liées à ces nouvelles lutheries. Dans le passage de l'instrument mécanique à l'instrument électrique, électronique et, plus encore, informatique, s'est opérée une rupture des chaînes causales traditionnelles qui garantissaient la cohérence temporelle, spatiale et énergétique des interrelations geste-instrument-son. Les études portant sur cette thématique ont connu ces dernières années un essor considérable.

**Etat de l'art en Linguistique Cognitive**

Au sein des sciences du langage, la linguistique cognitive s'attache à expliciter les relations entre langue, discours, cognition et pratiques [13, 18, 36] en étudiant les ressources linguistiques et les processus discursifs mis en œuvre par les individus pour décrire et partager leur expérience sensible de l'environnement [17,19]. Plus précisément ces ressources et processus participent à la construction et la structuration de notre expérience du monde dont il s'agit de rendre compte dans une analyse nécessairement pluridisciplinaire impliquant sciences physiques et sciences humaines et de la culture. Dans ce cadre, de précédentes recherches menées au LCPE - LAM se sont par exemple intéressées aux discours d'évaluation de la qualité des pianos [12], à l'évaluation de différents types de restauration d'enregistrements anciens [33], et à la notion de timbre précisément liée à la qualité de l'instrument [10] et différemment évaluée selon les sujets ayant différentes connaissances ou pratiques des instruments [2].

**Objectifs**

Il s'agit ici de mettre à contribution des outils conceptuels et méthodologiques développés en linguistique cognitive pour tenter de mieux comprendre la notion d'instrument en Informatique Musicale, à travers l'analyse de différents types de corpus. À la différence des travaux existants qui s'intéressent principalement aux caractéristiques physiques ou logicielles des dispositifs de MAO[1], ce travail vise à appréhender le concept d'instrument que ces nouvelles pratiques musicales induisent et que les utilisateurs convoquent et re-construisent dans leurs discours. Cette étude est également conçue comme un préalable nécessaire à la mise en œuvre et au développement de protocoles d'évaluation de nouveaux dispositifs musicaux, là où des méthodologies plus adaptées aux situations de laboratoire font défaut.

**Contribution principale**

Une analyse contrastive de différentes définitions de la notion d'instrument relevant i) du domaine lexicographique, ii) du domaine musicologique, iii) de pratiques terminologiques (implicites) dans la littérature en Informatique Musicale, ainsi que iv) de discours recueillis au cours de questionnaires et d'entretiens auprès d'utilisateurs (créateurs, développeurs, musiciens compositeurs et/ou interprètes, chercheurs, pédagogues, élèves) de dispositifs de MAO a été effectuée. La mise en perspective de ces différents types de discours à propos de l'instrument permet de préciser les différentes valeurs sémantiques que revêt ce concept selon les pratiques et les discours dans lesquels il s'inscrit [20]. Ainsi par exemple les dictionnaires définissent l'*instrument* différentiellement par rapport à l'*outil* tandis que les publications en Informatique Musicale traduisent une ambivalence entre la notion d'*instrument* et celle d'*interface*. La description de dispositifs de MAO par différents utilisateurs témoigne également d'un concept d'instrument aux frontières floues, les dispositifs étant considérés comme appartenant plus ou moins à la catégorie instrument, selon un « air de famille » [16].

**Retombées**

- Sur un plan théorique, cette étude permet de mieux saisir les enjeux et les conséquences du changement de paradigme à l'œuvre dans les Nouvelles Lutheries, en requestionnant un concept souvent considéré comme ontologiquement fondé et déplaçant alors la notion d'instrument comme entité à celle d'un objet défini quant à ses fonctionnalités lui permettant d'assurer la musicalité de ses productions sonores.
- Du point de vue de la création et de la pédagogie musicale, les réflexions ainsi menées peuvent aider à la réalisation de dispositifs musicaux mieux acceptés en tant qu'instruments.
- Enfin, ce travail, en amont d'une expérience d'évaluation de dispositifs de MAO (projet ANR Riam 2PIM) [8], permet de préciser le rapport des utilisateurs à ces objets et ainsi de contribuer à la mise en place de protocoles d'évaluation écologiquement valides [24], ce au-delà du domaine musical dans lequel s'inscrit cette recherche.

---

[1] Musique Assistée par Ordinateur

## 1. INTRODUCTION

As far as music is concerned, instruments have always been part of a cultural "landscape" (on technical, expressive and symbolic levels). The present contribution explores the changes brought about by the shift that occurred during the 20th century, from mechanical to digital instruments (also named "virtual instruments").

First and foremost, a short recall of some historical steps of the technological developments that have renewed our relationship to sound, music, and instruments will be presented. Second, an analysis of different discourses and terminologies presently used in the domains of musicology and computer music will account for the evolution of the notion of instrumentality.

## 2. FROM MECHANICAL TO DIGITAL INSTRUMENTS

History of music and its instruments is not independent of history in general as far as humans are concerned: history of ideas, history of arts, history of technology, history of their languages, aesthetical and emotional values.

### 2.1. The Promise of Ubiquity

The invention of the telephone and the phonograph, almost contemporary to each other, disturbed our "traditional" relationship to sound, voice and music, by allowing transmission and recording of sounds. From then, sounds could "travel" through space and time. Artists and poets, such as Villiers de L'Isle-Adam, Marcel Proust, Guillaume Apollinaire, among others, have been very impressed with these inventions and their consequences.

Developments in electricity and digital technology later increased the "distance" between the body and the instrument in music production:
- electricity, by bringing new energy, previously mechanical, to machinery and instrumental devices;
- digital technology by operating a radical decoupling due to symbolic encoding.

The consequences of these decoupling on musician-instrument interactions have been studied extensively by Cadoz [4,5] in particular. Although we will not repeat these discussions here, it is crucial to remind that these technologies induced a new definition of social value: from a world valorizing labor to a world based on information. Such evolution has artistic, cultural and social irreversible consequences. Therefore, it is not surprising that, alongside with these developments, new artistic sensibilities, new philosophical and scientific paradigms have been invented and experienced.

### 2.2. Mourning the Lost Unity

One consequence of these transformations can be described using the metaphor of Ancient Greek drama: the rule of the three unities (unities of Time, Place and Action) can be transferred to some extent to the "classical" relationship between listeners and sound phenomena. However, as we have seen, these three unities, which ensured the consistency of the instrumental relationship, have been disrupted by the development of new technologies.

Consequently, the musician and the listener, who used to be directly related to primary sound sources, are now, due to new artefacts, confronted to secondary sources producing *simulacra* without any guarantee of coherence.

### 2.3. New instruments: News definitions

Because of these decoupling, new questions and new musical devices emerged at the same time.

What is an instrument? What makes an instrument an instrument? How does an interface get its "instrumentality"?

Usually, most of research on this topic focuses on physical characteristics of devices, on the software underlying them, or on the sensorymotor interaction between the musician and the "interface". As a counterpart, it can be useful and productive to work on a human-centred approach taking into account the cultural and social aspects of the interactions between subjects and these new devices.

## 3. DEFINING INSTRUMENTALITY: CONTRIBUTION OF LINGUISTICS

The work presented in this paper is part of a collaborative project[2] about some specific computer music devices - Meta-Instrument [26] and Meta-Mallette [27]. Our main contribution aims at exploring different discourses and practices related to these devices, and getting at a better understanding of the new interactions between users and them. It could help designing new protocols of evaluation of musical devices, and eventually contribute to their improvement. Therefore, a new field investigation has been carried out exploring different kinds of discourses in the domain of computer music, in order to study how these new kinds of musical interactions, as practices, help us define a new kind of instrumentality. It will be done by analysing how people talk about it and therefore, through their discourse, contribute to a dynamic negotiation of new definitions and conceptualisations.

---

[2] ANR-2PIM Project, involving music associations and laboratories (Puce Muse, Grande Fabrique, Labri, LAM, LIMSI, IRCAM, McGILL).

Such a goal can be achieved using conceptual and methodological frameworks developed in human sciences, and more specifically in cognitive linguistics.

### 3.1. Cognitive Linguistics Framework

Individuals can provide and share across their discourses their sensory experiences related to the environment. Cognitive linguistics study relationships between language, discourse, cognition and practices by analysing linguistic resources and their organisation in discourse [13, 18, 36, 37]. These verbal resources and discourse processes contribute to the construction and the structuration of one's world experience [17, 19].

Within this pluridisciplinary approach, involving humanities and physical sciences, previous investigations have been carried out at LAM-LCPE concerned with different types of evaluative discourses on musical (piano or voice quality [11,33]) and non musical sounds (such as urban soundscapes [23]), as well as on other sensory modality (such as visual spaces [6]). To our knowledge, this kind of work has not been done yet in computer music except by Stowell et al. [41] who analysed verbal descriptions produced on new interfaces people were experiencing, in English language and culture and without any specific linguistic analysis. Our work brings new insights in this field of investigation: as an exploratory inquiry, it remains a prerequisite before designing new evaluation protocols for new musical devices.

### 3.2. General Methodology

Various kinds of discourses dealing with instrument and instrumentality have been collected:
- Dictionary definitions in French and English (§ 4),
- Current definitions and discussions about instrument in musicology, ethnomusicology and organology (§ 5),
- Terminologies used in computer music literature examining paper titles of recent publications in French and English (§ 6),
- Definitions collected through a written questionnaire addressed to members of the French computer music community during its annual conference: JIM09 (§ 7),
- Definitions and discourses in French about instrument and instrumentality given by different kinds of users during an interview dealing with their practices of computer music (§ 8).

Most part of this linguistic work has been done in French so as to compare lexicographic and terminological discourses with the spontaneous productions of the users (among the French community of computer music). Nevertheless, for the purpose of this article, we shall provide some brief analysis of English terminologies and definitions that will allow us to highlight some points of comparison, bringing also more evidence on the crucial role of language diversity in conceptualisation. All French citations are translated and presented in English in the body of the text, while original French cotes are presented in footnotes.

The contrastive analysis of these different discourses about instrument will allow us to specify the variety of meanings this concept can endorse depending on discourses and practices it is involved in.

### 4. *"INSTRUMENT"* IN DICTIONARIES

*Instrument* definitions from two French dictionaries: le Petit Robert (PR) [38] and le Trésor de la Langue Française (TLF) [28] were first compared.

PR defines *instrument* as a "manufactured object that is used to execute something, to do some operation"[3] and specifies "*instrument* is more general and less concrete than *tool* and refers to simpler objects than *apparatus* or *machine*"[4]. After exemplifying some kinds of instrument, a second entry is dedicated to musical instruments through a typological enumeration based on an organological classification (*areophon, cordophon, (...) instruments of the orchestra ...*), without any definition.

TLF defines *instrument* as "a concrete thing allowing to act on the physical world"[5] and makes a distinction between:
- a **generic signification** : "object made for a particular use, in order to do or create something, to execute or facilitate an operation (in technique, art or science)"[6];
- and a **specific signification in music** : "object entirely made or prepared from another natural or artificial object, the former being conceived to produce sounds and to serve as expressive means for composers and performers"[7].

First, it has to be noted that PR proposes an **opposition between *instrument* and *tool***, that a French author Simondon, specialist of the epistemology of technique, emphasized by defining the tool as a prolongation of the body to accomplish a gesture, and the instrument as a prolongation of the body to get a better perception[8] [40]. This perceptive aspect specific to the instrument is interesting even though recent works in psychology and neurophysiology inspired by phenomenology have recently reconsidered the perception action coupling [43].

---

[3] "Objet fabriqué servant à exécuter quelque chose, à faire une opération."
[4] "*Instrument* est plus général et moins concret que *outil* et désigne des objets plus simples que *appareil*, *machine*."
[5] "désignant une chose concrète permettant d'agir sur le monde physique."
[6] "Objet fabriqué en vue d'une utilisation particulière pour faire ou créer quelque chose, pour exécuter ou favoriser une opération (dans une technique, un art, une science)."
[7] "Objet entièrement construit ou préparé à partir d'un autre objet naturel ou artificiel, conçu pour produire des sons et servir de moyen d'expression au compositeur et à l'interprète".

As for TLF, it insists on the notion of **creation** and specifies domains in which the instrument is involved, namely **technique** as well as **science** and **art**. Moreover, TLF gives a specific definition of **musical instrument** as an **object producing sounds** with a **focus on expressivity and users** (composer and performer) of these objects.

Examination of two definitions of *instrument* in English dictionaries (Webster's New World Dictionnary – WD, [34] and New Oxford American Dictionnary – OD, [35]) leads us to clarify these specificities but also shows some contrasts between languages that may explain differences in conceptualisations. The instrument is defined as a "specific tool used for specific purposes (scientific or artistic) and delicate work", insisting on the same aspects developed in French by PR (tool/instrument) and TLF (domain specificity). Musical instrument is then defined as "an object or device for producing musical sounds" but **without any explicit mention of agency** as in TLF definition.

This first collection and analysis of *instrument* definitions shows different points of view, depending on what domain and level is taken into consideration (instrument *vs* tool, musical instrument). **The instrument then does not appear as a unique concept**.

### 5. "INSTRUMENT" IN MUSICOLOGY

Among others, ethnomusicologists have to deal with the diversity of human cultures. They may be considered as being at the right place to teach us how it is possible to categorize of musical instruments by describing their mechanical features, as organology does it. But the situations and symbolic systems they are used in and furthermore, part of, have to be taking in account. Thus, Michaud-Pradeilles [32] explained that:

> *"In order to define what is a musical instrument, one would better study the object by focusing on its function or its use without any restrictive criterion, such as the fact that it is, or not, made by a human being. If so, the useless distinction between musical sounds and noise could be avoided."[9]*

Within different words, Schaeffner, in the introduction of his book about "*The Origin of Musical Instruments*" [39], characterizes as a vain attempt any ontological definition:

> *"Can we define the term musical instrument? It is impossible, as well as we cannot state any precise definition of music that would be valid in every situation, every period, and every use of this art. The problem of instruments rejoins the question of boundaries of music. An object can be sonorous; how and why can we say it is musical? For which kind of qualities, Music will promote it to the same grade as others instruments?"[10]*

In addition to the difficulty to defining what is music, and then, what is a musical instrument, ethnomusicologists, such as Dournon [15], remind us that we cannot reduce instruments to their capability to produce sounds, such a definition being, indeed, unable to fit the diversity of acoustical situations and social contexts:

> *"A musical instrument is not an object as others; it produces sounds and it also brings meaning. It includes an additional aspect, due to its functional and symbolic role in society."[11]*

### 6. TERMINOLOGIES IN COMPUTER MUSIC

In computer music literature, a broader diversity of designations and denominations referring to hardware and software objects can be found in papers and their titles. A non-exhaustive list of these designations, mainly collected in the proceedings of past conferences (NIME, SMC, JIM), in English and in French reveals a rich lexical, syntactic and semantic diversity, that can be structured by two mains lexical forms, *instrument* and *interface* (see table below).

---

[8] "Le XVIII$^e$ siècle a été le grand moment du développement des outils et des instruments, si l'on entend par outil l'objet technique qui permet de prolonger et d'armer le corps pour accomplir un geste, et par instrument l'objet technique qui permet de prolonger et d'adapter le corps pour obtenir une meilleure perception ; l'instrument est outil de perception. Certains objets techniques sont à la fois des outils et des instruments, mais on peut les dénommer outils ou instruments selon la prédominance de la fonction active ou de la fonction perceptive." ( [40], p.114)

[9] "Pour définir l'instrument de musique, il vaudrait mieux considérer peut-être l'objet par rapport à son rôle ou à l'usage qui en est fait sans apporter de notion restrictive, telle que la participation de l'homme quant à son élaboration et éviter la ségrégation d'ailleurs magistralement remise en cause de nos jours entre sons musicaux et bruits." ([32], p. 5)

[10] "Pouvons-nous définir le terme d'instrument de musique ? Autant peut-être nous demander s'il existera jamais une définition de la musique, qui soit précise et valable en tous les cas, qui répondent également à toutes les époques et à tous les usages de cet art. Le problème des instruments ne touche-t-il pas à celui des limites de la musique ? Un objet est sonore ; à quoi reconnaîtrons-nous qu'il est musical ? Pour quelles sortes de qualités la musique le mettra-t-elle au rang de ses autres instruments ? " ([39], p. 9)

[11] "L'instrument de musique n'est pas un objet comme les autres, il est un outil à la fois producteur de sons et porteur de sens. Il comporte en effet une dimension supplémentaire déterminée par le rôle fonctionnel et symbolique qu'il joue dans la société. " [15]

| | French designations | English designations |
|---|---|---|
| INSTRUMENT | *instrument numérique* | *digital instrument* |
| | | *digital music instrument* |
| | *instrument logiciel* | tr. : software instrument |
| | *instrument virtuel* | *virtual instrument* |
| | | *sound sustain virtual instruments* |
| | | *virtual and tangible instrument* |
| | | *interactive instrument* |
| INTERFACE | *interface musicale* | *musical interface* |
| | | *digital musical interface* |
| | | *expressive musical interface* |
| | | *computer music interface* |
| | | *new interface for musical expression* |
| | | *tangible acoustic interface* |
| | *interface de communication instrumentale* | tr.: instrumental communication interface |
| | | *from musical interface to musical instrument* |

**Table 1**. Interface and instrument denominations in computer music litterature

Legend: all designations in italic are original titles and citations. When designations appeared in both langagues, they have been paired, otherwise a translation is proposed (tr.: without italic).

### 6.1. Musical Interfaces & Digital Instruments

In this specific area, *instrument* is no longer related to *tool* but now to *interface*[12]. This last term is both generally defined in English and French[13] as "a point where two systems or subjects (…) meet and interact" and mainly in physics as "a surface forming a common boundary between two portions of matter or space". A specific definition in the field of computing is given by OD (and TLF in French): "a device or program for connecting two items of hardware or software so that they can be jointly operated or communicate with each other".

On the one hand, *interface* is more generic than *instrument* and can explain why the attributive adjective *musical* is used in titles to specify the area to which it applies (ex: *musical interface*). On the other hand, its use specifies the relation to the technological domain and puts emphasis on the **relationship between human and computer**. This argument can be used to explain the scarcity of the denominations combining *interface* and adjectives such as *virtual* or *digital* in the corpus, as well as the frequency of denominations including a combination of *instrument* and the same adjectives. It would be a little pleonastic insofar as *interface* already includes a computing aspect. Likewise, there is no need either for additional adjective *musical* to qualify *instrument* in these titles. Considering linguistic constructions, a double dynamic (movement) can be indeed identified as illustrated in the figure below:

---

[12] Others denominations such as *interactive installation*, *controllers*, *augmented instrument*, *interaction* can also be found more rarely.
[13] Considering that *interface* is in French a loanword from English, we just mention English definition in this paper.

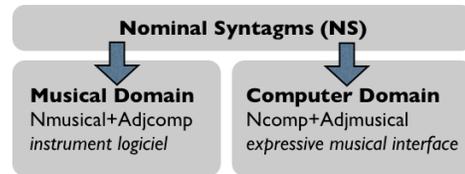

**Figure 1**. Construction of Nominal Syntagms (NS) in computer music titles

Within the musical domain, *instrument* as a noun will be associated with (/ specified by) an adjective referring to computing whereas within the computer domain, the noun interface will be associated with an adjective that will confer to it a musical aspect. Though, depending on the domain of reference chosen, emphasis would be more on music or on computing.

*Interface* also introduces an **uncertainty about the materiality** of the referred object. It becomes apparent that a semantic shift occurred. In dictionaries *interface* puts together two matters, objects, systems, individuals i.e. two "things" that are considered at the same level. Titles collected here point out another type of interaction: between human and computer. Then the question becomes: is *interface* referring to **physical interface** between musician and computer (as *tangible* seems to illustrate) or to **software interface** between musician and computer hardware?

As already shown, *instrument* denominations are mostly constructed with *instrument* as head noun of nominal syntagm qualified by attributive adjective(s) referring to computer: *numérique / digital, logiciel / software, virtuel / virtual* … The issue of "virtuality" in computer music and more generally in human computer interaction studies is very important and would deserve a longer development (for discussion and extension to visual domain, see all also [7, 25], and more generally on Virtual Reality see [29]). We just want to focus here that all denominations based on *instrument* (or on *interface*) emphasize more on digital aspects than on physical devices with which they are coupled.

### 6.2. Instrumental Interfaces

Nonetheless, one French denomination differs from the others by joining *instrument* and *interface*: *interface de communication instrumentale* (tr: instrumental communication interface). In this title, *interface* is qualified by the adjective *instrumental* (built on noun *instrument* + suffix–*al*). The device is then referred to as an **interface characterised by its instrumentality**. This denomination from Cadoz [4] has to be linked with another designation (in English) this author and his collaborators use: *instrumental* (*gesture*) *interaction* [5]. The fact that *instrumental* is preferred to *instrument* in their terminology indicates a paradigm shift. Understanding and describing interaction dynamics

(mostly in gesture) between the musician and the physical and digital devices he or she uses allow evaluating the instrumentality of these devices. In this approach *instrumental* is defined[14] as:

> "Any relational activity which needs a material device (instrument) exterior to human body, such as it needs an energy consumption outside of the frontiers of the body and of the instrument, and at least one part of the produced energy comes from the human body"[15].

In short it can be considered as a **shift from *instrument* as an entity to *instrumental* quality**[16].

In examining the diversity of denominations and conceptualisations in computer music terminology, it can be concluded that the innovative character of this domain implies a first stage of proposition, confrontation and negotiation with different designations before reaching an accepted, stabilised and consensual terminology [3, 14].

However, the recurrent lack of explicit definitions and the few research dedicated to the definition of the concepts used are associated with the paradigm shift these new musical technologies induce. When computer music instrumentality is questioned, investigations are more often centred on devices themselves. Implicitly instrumentality seems to be coupled with technological quality of the interfaces considered.

Nevertheless some studies focus on musician-interface sensorymotor loop [4]. Introducing this first level of interaction, perception-action coupling is taken into account. However, a further advancement in this dynamical movement between interface (object) and musician (subject) has to be proposed. The analysis of how experts define what is an instrument and how users talk about the devices they use and their relation to them will constitute an attempt to fulfil this requirement.

## 7. EXPERTS' DEFINITIONS OF INSTRUMENT

During last JIM Conference (annual conference of the French association of computer music) that took place in Grenoble in April 2009, a paper discussing some first results of this research has been presented [9]. This meeting was an opportunity to ask participants to define what is an instrument (*In your opinion, what is an instrument?*[17]). 30 definitions[18] were thus collected.

Three domains of definition could be identified (general, musical/sonic and computer music) as illustrated in table 2.

| Domain | Nb of def. | Examples |
|---|---|---|
| General | 15 | - instrument, it's an extension of the human body[19]<br>- an instrument is a material object which will be useful for us in a concrete goal[20] |
| Sonic Musical | 10 | - device that we manipulate to produce some sound[21]<br>- that is a tool that allows the musical expression[22] |
| General + Music | 3 | - the instrument could be an interface between the human and the task one wants to perform (ex: sound production)[23] |
| Computer Music | 1 | - It is one or some controllers that the musician can manipulate and that influences some musical parameters. It can also be a software[24] |

**Table 2**. Domains of definition of *instrument*

Most of the collected definitions take place in a general domain, or concern directly music and sonic domain. A few others have a double structure, beginning first with a general concern and going then continuing with a specific part concerning music or sound.

In these sentences, people rather define *instrument* as a **tool**, sometimes as an *object*, and less often as a *device* or an *implement*[25]. The instrument is also often described as an **extension**/*prolongation* of the human body and as an *interface - intermediary* between the human and the world. Most of questioned people mentioned that it is something that **allows** to do something, something *for* ...

Paying attention to the presence of the speakers in their definitions, different strategies can be observed going from a maximised objectivation (*instrument = transformer*[26]) to the explicit relationship between the subject, the instrument and the world (*any object I can use to something, to represent an idea upon different aspects (sound, image, construction, other object*[27]).

---

[14] This is one of the few works in which these notions are explicitly defined and specified.
[15] "[on peut définir] comme instrumentale toute activité relationnelle qui fait appel à un dispositif matériel (l'instrument) extérieur au corps humain et telle que sa finalité nécessite une consommation d'énergie en dehors des frontières du corps humain et de l'instrument ; et une partie au moins de cette énergie provient du corps humain." ([4] p.61)
[16] This dynamic is also illustrated in another NIME publication: *T-Stick : from musical interface to musical instrument* [30].
[17] *Selon vous, qu'est-ce qu'un instrument ?*
[18] including 29 written definitions and one diagram.
[19] *Instrument, c'est une extension du corps humain*
[20] *Un instrument est un objet matériel qui va nous servir dans un but concret (...)*
[21] *Appareil que l'on manipule pour produire du son*
[22] *C'est un outil qui permet l'expression musicale*
[23] *l'instrument pourrait être une interface entre l'homme et la tâche que l'on souhaite accomplir (ex : production de son)*
[24] *C'est un ou des contrôleurs que le musicien peut manipuler et qui influe sur des paramètres musicaux. Ça peut être aussi un programme*
[25] *outil, objet, dispositif, ustensile.*
[26] *Instrument = transformateur*
[27] *tout objet que je peux utiliser pour façonner quelque chose, pour représenter une idée sous différentes formes (son, image, construction, autre objet ...)*

Eventually, some answers insist on playing and on the expressivity that are both characteristic features of music:

> *An instrument is something which one knows how to play. That addresses the question: what is "to play" ? It is to make music so that it is heard by the listener*[28]

These last considerations underline the importance of the instrumental relationship (playing music) between the musician and the object, and also between the musician and the listener. Considering that, we know focus on the way users of these new digital musical devices describe their practices and the devices they are used to play with.

## 8. INSTRUMENT IN USERS' DISCOURSE

### 8.1. Presentation of the Study

As already described in section 3, this work contributes to the 2PIM project and relies on studying practices and discourses of computer music users concerning specific devices: Meta-Instrument [26] (see figure 2) and Meta-Mallette [27] (see figure 3).

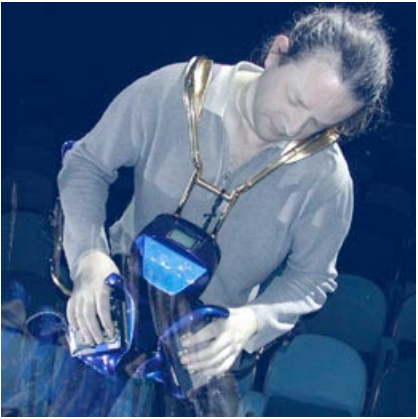

**Figure 2**. S. de Laubier, Meta-Instrument creator

Developed by the group Puce Muse, they are both generally described as controllers mapped with software processing sound and visual synthesis. Whereas Meta-Instrument is an ad hoc device preferentially assigned to individual and expert musical practice, Meta-Mallette uses commercial interfaces (joystick essentially) and is conceived to be played immediately (without any necessary prerequisite) and collectively (joysticks orchestra).

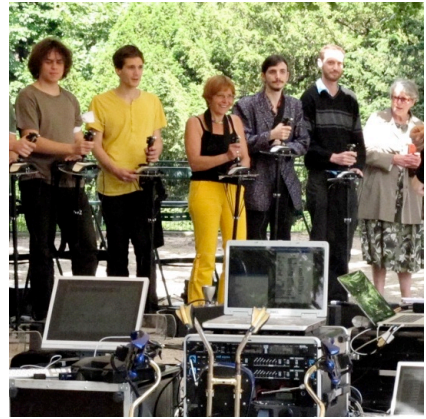

**Figure 3**. the Meta-Orchestra playing Meta-Mallette

10 users of these devices (creators, developers, composers and/or performers, researchers, teachers, pupils from conservatoire)[29] have been interviewed using semi-directive methodology. Subjects were asked non-directive questions (ex: *in your opinion what is Meta-Instrument? how do you characterise it?*[30]), going from the more general to the more specific thematic[31]. At the end of interviews they were also asked to answer to an open questionnaire concerning some words: sound (*son*), visual form (*forme visuelle*), gesture (*geste*) and instrument (ex: *could you give 5 examples of instrument? In your opinion what is an instrument?*[32])

This questionnaire has various objectives:
- It allows us to obtain comparable discourses of users;
- It mainly enables to contrast definitions that emerged during interviews and those given as answers to this questionnaire. In the first two interviews, subjects were only asked to define *son*, *forme visuelle* and *geste*. Then, during these interviews, it appeared necessary to also question them about *instrument*[33].

After an integral transcription, linguistic analyses have been conducted at different levels, respectively lexical, morphosyntactic, semantic and discursive. This allowed us to draw inferences about relationships between speakers, the instruments they were playing and/or developing and audiovisual musical production

---

[28] *Un instrument, c'est quelque chose dont on sait "jouer". Ce qui pose la question : qu'est-ce que "jouer" ? C'est faire de la musique pour qu'elle soit entendue par l'auditeur.*

[29] All people interviewed have already used Meta-Instrument and/or Meta-Mallette. Some of them combine different competencies and also use other devices they have or not have created.

[30] *Pour toi c'est quoi le Méta-Instrument ? Qu'est-ce qui le caractérise ?*

[31] The questions can be quite different depending on people interviewed, unless some topics were always addressed.

[32] *Peux-tu donner 5 exemples d'instrument ? Pour toi qu'est-ce qu'un instrument ?*

[33] This questionnaire contributes to a research program pursued at LCPE on sensory objects such as colour, sound, odour, gesture, etc. [17].

generated during these interactions. Some results are presented in the next section.

## 8.2. Results

### 8.2.1. What is an instrument? Examples of definitions

Interviewees provided at the end of the interviews different types of definitions of *instrument*. Some of them first define a **generic instrument**[34]:
- *something that is used to instrument (...) an instrument it's used to transform the world* (S3)
- *it's the link between doing something and the will to do it* (S5)

before talking about music in a second part of their definition (except S3).

Others directly refer to a **musical instrument**[35]:
- *it's .. a ... a tool of audio rendering yes that transcribe a gesture or a thought into sound* (S6)
- *an instrument well it's a it's a tool that allows to make sounds* (S7)

Definitions are be structured around:
- instrument **properties**[36]:
  - *it's a mixture of stuff that has been thought for a very precise action* (S9)
- what instrument **allows to do**[37]:
  - *to transform the world* (S3),
  - *to make sounds* (S7),
  - *possibilities to touch the sound* (S8)
- what instrument is **characteristic of**[38]:
  - *this is what makes us human* (S5)

considering that a same definition could contain several of these different aspects.

A part from these definitions centred on instrument as an entity, some speakers shift the focus **from the instrument to the instrumental function**[39]:
- *we can give an instrumental function to things that are not instruments* (S4)

This plurality of instrument conceptualisations is crucially revealed when we pay attention to specific devices.

### 8.2.2. Instrumentality: a polymorphous concept

Considering the specific devices this study investigates, the questions about instrument examples and definitions at the end of interviews lead to two main results:
- Meta-Instrument and Meta-Mallette are sometimes spontaneously given as **examples of instrument** (more frequently in case of Meta-Instrument)
- Defining the instrument leads users to question themselves upon the **instrumental nature of devices** they have been discussing during their interview.

Definitions are built on **successive oppositions** between Meta-Instrument and Meta-Mallette. This can be illustrated by the following extract from S1 opposing
- **a meta-INSTRUMENT**:
  *what is it, in your opinion, the Meta-Instrument? well as I told you for me it's an instrument that I rather refer to as a musical instrument*[40]
- and **a meta-DEVICE** (in french : *méta-dispositif*)
  *so as for the Meta-Mallette, unlike the Meta-Instrument, I would say that it's less of a musical instrument and more of a musical and visual device*[41]

Regular marks of **negotiation concerning boundaries of** what is considered as an **instrument** are frequently produced. If most of the interviewees give an instrumental status to the Meta-Instrument, nevertheless; depending on the context, they spend time successively excluding and including the Meta-Instrument from the instrument category. In short it can be at the same time:
- **less than an instrument**: it needs a software (also called *software* or *virtual instruments*) and a computer:
  *the object in itself ... I wouldn't put it in the instrument category ... it's the Meta-Instrument plus e:: what we called the software instrument (...) it's the couple made up of Meta-Instrument and the interpretation algorithm* (S6)[42]
- **more than an instrument**: compared to other instruments, its necessary connection to computer provides additional properties that transform it into *super*, *hyper* or *meta* (!) instrument:
  *yes I think they are instruments, it's more than a simply instrument because in fact inside they have a memory (...) because even Serge's instrument has a memory too* (S9)[43]

---

[34] - *quelque chose qui sert à instrumenter (...) un instrument ça sert à transformer le monde*
- *c'est le lien entre faire quelque chose et la volonté de le faire*
[35] *c'est ... un ... un outil de de rendu (...) audio ouais qui transcrit que ce soit un geste ou une pensée en son*
*un instrument ben c'est un un outil qui permet de faire des sons*
[36] *un instrument c'est un assemblage de bidules qui a été pensé (...) pour une action euh très précise*
[37] *transformer le monde, faire des sons, des possibilités de toucher le son*
[38] *un instrument c'est ce qui fait de nous des êtres humains*
[39] *on peut euh donner une fonction instrumentale à des choses qui sont pas des instruments*

[40] *donc comme moi j'te disais pour moi donc c'est un instrument voilà donc ... que j'référence plutôt à un instrument d'musique.*
[41] *du coup la Méta-Mallette contrairement au Méta-Instrument j'dirais moins qu'c'est un instrument de musique mais beaucoup plus un dispositif musical et visuel (...).*
[42] *non en lui-même je le mettrais pas dans les instruments c'est le méta instrument plus euh ce qu'on appelle l'instrument logiciel (...) ça serait plus le couple Méta-Instrument et algorithme d'interprétation (...)*
[43] *je pense oui que ce sont des instruments c'est plus qu'un instrument parce que en fait dedans ils ont des mémoires (...) parce que même l'instrument de Serge aussi il a une mémoire*

Moreover, if the **Meta-Mallette is referred to as a device** by most users, like S1:

> *so as for the Meta-Mallette, unlike the Meta-Instrument, I would say that it's less of a musical instrument and more of* a musical and visual device made up of a software and of instruments ... *which* are joysticks *for the time being*[44]

It can be noticed that in the context of a collective musical practice **S1 calls the joysticks *instruments***. About this, S1 explains:

> *In fact when I'm talking about it I gonna say* instrument for joystick *meaning* meaning you take your instrument to play it *(...) so you gonna take the object joystick to play (...) and in the same time commonly when one talks MetaMallette when one knows a little what's in (...) when* we talk about instrument this is the small software part *that determines how your sounds and images gonna react*[45]

Thus, the **joystick is not *an instrument* but becomes *the user's instrument*(s)** as soon as she/he is playing, mainly as a participant of an orchestra as in the example presented (for other digital orchestra experiment see also [41]). In short, once involved into musical practices, especially collective ones (with history, cultural and social values) devices can be considered as instruments. Then, while an "objective" definition (i.e. that aims to abstract the defined object from any contexts) leaving out the musician, can deprive the device from any instrumentality, as soon as a user is considered and makes it his / hers, it can more easily acquire instrumentality. Thus, as an another interviewee explains:

> *But for me it is, the musical instrument it is the one which vibrates (...) the MetaMallette I believe that ... as the others instruments (...) I put it in the instrument category because I think one can be able to make it vibrate*[46] – S5

Furthermore, it is interesting to notice that expert users (of the Meta-Instrument) say they make *their instruments*, talking about software they have developed. This introduces another important point concerning expert know-how and knowledge in designing new digital musical devices, called in French *nouvelles lutheries* or *lutheries electroniques*.

## 9. CONCLUSIONS & PERSPECTIVES

Considering computer music practices through linguistic analyses of discourses, it appears that "*instrument*" does not actually refer to a device (hardware and software) but rather qualifies its interaction with users (musician, designer, etc.). Thereby it links the instrumentality to the emergence of an "intangible" repertory of gestures, situations, repertoires, possibly shared. Eventually, it points out that this immaterial part is at least as important as the physical one.

The instrumentality of these new devices, as well as of "classical" instruments, does not result from their intrinsic properties only. It is constructed through musical play, interactions between musicians and the design and development of the instruments, as illustrated by one user's comment: *one is not born, but rather becomes, an instrument*[47] (S4). This evolution and enrichment of the concept of instrumentality that these new practices make re-emerge impose a pluridisciplinary approach for the "science of instruments". Such considerations have already been pointed out in ethnomusicology, as illustrated by the fieldwork of Marcel-Dubois [31] (in Gétreau's dissertation [22]):

> *Claudie Marcel-Dubois* (that largely contributed to the knowledge of traditional instruments) *has achieved her career in 1980 with the biggest temporary exhibition of traditional instruments ever realized in France. Called "The popular musical instrument. Uses and symbols", she assigned a sort of manifest: the instrument was presented there considering its ritual and symbolic functions, its mobility (from popular to scholarly and reciprocally), its geographic variations, and its morphologic, semantic and classificatory components"*[48]

Furthermore those theoretical and epistemological considerations have already contributed to the development of new protocols for evaluating new digital instruments, accounting for their practices and cultural values [8].

---

[44] *(…) du coup la Méta-Mallette contrairement au Méta-Instrument j'dirais moins qu'c'est un instrument de musique mais beaucoup plus un dispositif musical et visuel (...) qui va se composer d'un logiciel et d'un nombre d'instruments euh… qui pour le moment sont plutôt des joysticks*

[45] *En fait quand j'en parle je vais dire instrument pour joystick dans le sens ou … dans le sens où tu prends ton instrument pour en jouer (...) donc tu vas prendre l'objet joystick pour jouer (..) et en même temps communément quand on parle la Méta-Mallette quand on sait un peu ce qu'il y a dedans (...) quand on parle d'instrument c'est la petite partie logicielle qui va déterminer comment tes sons et tes images vont réagir*

[46] *Mais pour moi c'est, voilà l'instrument de musique c'est celui qui vibre.(...) la Méta-Mallette je crois qu'il … comme tous les autres instruments (...) je le mets dans les instruments parce que je pense qu'on peut arriver à le faire vibrer*

[47] *on ne naît pas instrument on le devient,* which mimics S. de Beauvoir's famous statement : "on ne naît pas femme, on le devient".

[48] *"Claudie Marcel-Dubois* (qui a largement contribué à la connaissance des instruments des instruments traditionnels, ndlr) *achève sa carrière en 1980 par la plus grande exposition temporaire qui leur fut jamais consacrée en France. Avec pour titre* L'instrument de musique populaire. Usages et symboles, *elle constitua une sorte de manifeste : l'instrument était présenté dans ses fonctions rituelles et symboliques, dans ses mobilités (du populaire au savant et inversement), dans ses variantes géographiques, dans ses composantes morphologiques, sémantiques et classificatoires"* ([22], p.21).

## 10. ACKNOWLEDGEMENTS

We thanks all partners involved in 2PIM ANR project, the JIM09 participants who accepted to answer our questionnaire and the people who accepted to be interviewed and to share their practices and knowledge. We also thank the readers who helped us with English translation.